\documentstyle[prd,preprint,aps]{revtex}

\begin{document}
\draft
\title{Thermodynamic approach to warm inflation}
\author{J. A. S. Lima \thanks{Electronic address: limajas@dfte.ufrn.br}
and
J. A. Espich\'an Carrillo\thanks{Electronic address:
espichan@dfte.ufrn.br}}
\address{Departamento de F\'{\i}sica, Universidade Federal do Rio
Grande do Norte, C.P. 1641 59072-970 Natal RN Brazil}
\maketitle

\begin{abstract}
We study the thermodynamic behavior of a decaying scalar field coupled
to a relativistic simple fluid. It is shown that if the decay products
are represented by a thermalized bath, its temperature evolution law
requires naturally a new phenomenological coupling term. This ``energy
loss'' term is the product between the  enthalpy density of the
thermalized bath and the decay  width of the scalar field. We also argue
that if the field $\phi$ decays ``adiabatically" some thermodynamic
properties of the fluid are preserved. In particular, for a field
decaying into photons, the radiation entropy production rate is
independent of the specific scalar field potential $V(\phi)$, and the
energy density $\rho$ and average number density of photons $n$ scale as
$\rho \sim T^{4}$ and $n \sim T^{3}$. To illustrate these results, a new
warm inflationary scenario with no slow roll is proposed.
\end{abstract}


In the new inflationary scenario accelerated expansion and reheating are
separated into two distinguished periods. The first one is an
exponential growth of the scale factor with the Universe evolving to a
supercooled state. Due to this adiabatic expansion the temperature of
the Universe decreases nearly $10^{28}$ orders of magnitude \cite{RB85}.
At the end of this supercooling process the Universe is reheated. The
field rapidly oscillates about the global minimum of its potential, and
the energy density of the inflaton field is completely or almost
completely converted into radiation in less than one expansion Hubble
time. As a matter of fact, either on its early \cite{Oldreheat} or
modern version based on parametric resonance (sometimes called
preheating) \cite{Newreheat,Kofman}, the reheating is a very fast and
extremely nonadiabatic mechanism.

In principle, if a sustained radiation component during inflation
is allowed, the supercooling and subsequent reheating could be
supressed or at least weakened by many orders of magnitude
\cite{lima88}. The first field motivated scenario based on this
idea is the isothermal or warm inflationary picture as proposed by
Berera \cite{B95}. Like in new inflation, the warm picture starts
from a high temperature phase transition with the universe
evolving through a de Sitter inflationary period dominated by the
scalar field potential. However, in the course of the expansion,
energy is continuously drained from the field $\phi$ to the
thermal bath. The whole process is described by an ``energy loss" term,
$\Gamma \dot\phi^{2}$. At the level of the scalar field equation of motion, it
contributes like an additional viscosity $\Gamma \dot\phi$, which may dominate
the term $3H\dot\phi$ corresponding to the redshift of $\dot\phi$ (the
momentum of field) by the expansion. In the warm picture, the persistent
thermal contact during inflation implies that the scalar field evolves in a
sort of over damped regime. As a result, the Universe approaches a state where
the dilution of the thermalized bath (due to expansion) may continuously be
compensated by the production of particles from the decaying scalar field,
thereby guaranteeing the constancy of the temperature. If this scenario works
during an interval of time long enough, thermal fluctuations may also
produce the primordial spectrum of density perturbations
\cite{Bererafang,Leefang}(see also \cite{B00} for an updated and more
detailed analysis).

More recently, this nice picture has severely been criticized in
terms of its physical plausibility by Yokoyama and Linde
\cite{YL99}. The basic argument is that $\Gamma \dot\phi^{2}$, is
not sufficiently strong in the regime where it should describe
warm inflation. This coupling seems to be only a small
contribution in a sub-leading thermal correction, and as such, it
is not expected to play a prominent role in the inflationary
process \cite{YL99}(see, however, Refs. \cite{B00} and
\cite{BGR}). On the other hand, as discussed in an extended
framework \cite{JL99}, the basic idea of warm inflation is so
attractive that it cannot be discarded without a more detailed
investigation or even further attempts to solve the above pointed
out difficulties. In particular, it is not so neat that an
``energy loss'' term like $\Gamma \dot\phi^{2}$ is a realistic
approximation for describing the energy dissipated by the $\phi$
field to a thermalized bath.

In what follows we apply basic thermodynamics arguments for studying the
decay  of the scalar field coupled to a thermalized bath which is
represented by a relativistic simple fluid. We recall that
thermodynamics has often been used when the underlying microphysics of a
given phenomenon has not been completely clarified. The basic reason is
that it yields relations among the macroscopic quantities whose validity
is independent of the microphysics on which they ultimately depend.
Therefore, it is interesting to consider some thermodynamic criteria on
the question related to the plausibility of warm inflation. As we shall
see, this macroscopic treatment clearly suggests a new form to the
coupling term, which depends explicitly on the created component through
its enthalpy density. It has a quite simple form, namely,
$\Gamma_{\phi}(\rho + p)$, where $\Gamma_{\phi}$ is the decay width of
the scalar field, and may alter significantly the studies of the
reheating period, as well as the original and extended warm inflationary
pictures\cite{N1}.

We will limit our analysis to homogeneous and isotropic FRW flat
universes. Following standard lines, the energy content is a mixture of
a coupled scalar field  plus a relativistic simple fluid representing
the thermalized bath. The total energy stress of this self-gravitating
mixture obeys the Einstein field equations (EFE). In addition, since the
scalar field works like a source of particles, the particle flux of the
relativistic bath ($N^{\mu} = n\,u^{\mu}$) must satisfy a balance
equation of the type $N^{\mu}_{;\mu}=\Psi$. The basic equations are:
\begin{equation}
\label{FR00}
{8\pi \over m^{2}_{pl}}\left({\dot\phi^2 \over 2} + V(\phi) +
\rho\right) \ = \ 3H^2,
\end{equation}
\begin{equation}
\label{FR11}
{8\pi \over m^{2}_{pl}}\left({\dot\phi^2 \over 2} - V(\phi) + p\right) \
= \ -\ 2\dot{H} - 3H^2,
\end{equation}
\begin{equation}
\label{w9}
\dot n  = -3nH + n\Gamma_{\phi},
\end{equation}
where the dot means time derivative, $H = {\dot R \over R}$ is the
Hubble parameter, $m^{2}_{pl} = 1/G$ is the Planck mass, $\rho$
and $p$ denotes the energy density and pressure, and $n$ is the
particle number density of the relativistic fluid which can be
radiation though is not restricted just to that. The first
quantity on the right-side (\ref{w9}) represent the dilution of
$n$ due to the expansion of the universe, while the second
describes phenomenologically the source of particles, that is, the
rate produced by the field $\phi$, which has been parameterized by
the decay width of the scalar field ($\Psi = n\Gamma_{\phi}$).
Note that (\ref{w9}) can be rewritten in more enlightening form
\begin{equation}
\label{w9a}
{dN \over dt} = \Gamma_{\phi}N
\end{equation}
where $N=nR^{3}$ is the number of fluid particles per comoving volume.
In particular, if $\Gamma_{\phi}$ is constant, the decaying of the
scalar field increases $N$ exponentially in the course of time. In the
the lack of a detailed model, $\Gamma_{\phi}$ is supposed to be a
generic function of the temperature. It should be noticed that
(\ref{w9}) and its possible dynamic influence has not previously been
considered in the warm context.  We argue here that such an approach is
thermodynamically inconsistent because (\ref{w9}) may change the
evolution of the universe even in a single fluid description with no
particle production \cite{CL89}, and, indeed, it is fundamental for a
consistent thermodynamic description of a decaying scalar field. Another
point is related to the nature of the interaction term. The energy
conservation law ($u_{{\mu}}{T_{t}^{\mu\nu}}_{;\nu}$), which is
contained in the EFE, may also be written as a balance equation for the
fluid
energy density
\begin{equation}
\label{w10a}
\dot \rho + 3H(\rho + p) = -\dot \rho_{\phi} - 3H(\rho_{\phi} +
p_{\phi} ),
\end{equation}
or equivalently,
\begin{equation}
\label{w10}
\dot \rho + 3H(\rho + p) = -u_{\mu}{T_{\phi}^{{\mu}{\nu}}}_{;\nu}.
\end{equation}
and may substitute the second FRW equation. In warm inflation
\cite{B95,TB20}, as well as in the first studies of reheating
\cite{MT83}, the above equation was solved assuming the phenomenological
coupling
$u_{\mu}{T_{\phi}^{{\mu}{\nu}}}_{;\nu}=-\Gamma {\dot {\phi}}^{2}$.
However, although very useful due to its simplicity, a physically
reasonable coupling term should be dependent on the nature of the
created particles and not only of the scalar field itself. These two
points are inextricably intertwined, in such a way that a
proper macroscopic treatment (taking into account (\ref{w9})) give rise
to an alternative coupling term.

The macroscopic quantities of the $\gamma$-fluid are related by the
Gibbs law
\begin{equation}
nTd\sigma \ = \ d\rho \ - \ {\rho \ + \ p \over n}\,dn,
\label{w13}
\end{equation}
where $\sigma$ is the specific entropy (per particle) and $T$ is the
temperature.

By adopting $T$ and $n$ as basic thermodynamic variables, one may show
from
(\ref{w9}), (\ref{w10})
and (\ref{w13}) that the evolution temperature law is
\begin{equation}
{ \dot T \over T}  =  -3\biggl({\partial p \over \partial
\rho}\biggr)_{_n}{\dot R \over R} -
\frac {n\Gamma_{\phi}\biggl({\partial \rho \over \partial
n}\biggr)_{_T}}
{T\biggl({\partial \rho \over \partial T}\biggr)_{_n}} -
\frac {u_{\mu}{T_{\phi}^{{\mu}{\nu}}}_{;\nu}}
{{T\biggl({\partial \rho \over \partial T}\biggr)_{_n}}}.
\label{w16}
\end{equation}
The first term on the RHS of (\ref{w16}) is the usual equilibrium
contribution. For an expanding fluid ($\dot R >0$), one finds $\dot
T<0$, as it should be. The remaining terms display the out of
equilibrium contributions due to the sources of particles
($n\Gamma_{\phi}$) and energy ($u_{\mu}
{T_{\phi}^{{\mu}{\nu}}}_{;\nu}$). A more convenient form to the
temperature law is obtained inserting (\ref{w9}) into (\ref{w16}) to
obtain
\begin{equation}
{\dot T \over T} \ = \ \biggl({\partial p \over \partial
\rho}\biggr)_{_n}\frac {\dot n}{n} \ - \
\frac {1}{T\biggl({\partial \rho \over \partial T}\biggr)_{_n}}
\biggl[(\rho + p)\Gamma_{\phi} \ + \
u_{\mu}{T_{\phi}^{{\mu}{\nu}}}_{;\nu}\biggr].
\label{w17}
\end{equation}
The interaction term ($u_{\mu}{T_{\phi}^{{\mu}{\nu}}}_{;\nu}$)
can be determined only if one imposes some thermodynamic constraint in
the
above expression. In order to do a realistic physical choice, we first
notice that the equilibrium thermodynamic relation for a radiation bath
($p={1 \over
3}\rho$), which is given by $n \propto T^{3}$, is possible only if the
second
term on the RHS of (\ref{w17}) vanishes identically, that is,
\begin{equation}
u_{\mu}{T_{\phi}^{{\mu}{\nu}}}_{;\nu} \ = \ -(\rho + p)\Gamma_{\phi}.
\label{w18}
\end{equation}
In general, for a fluid satisfying the $\gamma$-law equation of
state, $p=(\gamma - 1)\rho$, from (\ref{w17}) the particle number
scales as $n \propto T^{\frac{1}{\gamma-1}}$, as expected for a
thermalized $\gamma$-bath, only if (\ref{w18}) is satisfied. In
addition, replacing (\ref{w18}) into (\ref{w10}), and using
(\ref{w9}) one may verify that $\rho \propto
T^{\frac{\gamma}{\gamma -1}}$. Since the whole argument may be
inverted, it follows that the equilibrium relations for $\rho$ and
$n$ are preserved, thereby characterizing a thermalized bath, if
and only if the coupling term is proportional to the enthalpy
density of the created particles as given by (\ref{w18}). This
coupling term is completely different from $\Gamma
{\dot\phi}^{2}$, which has extensively been adopted in the
literature. As should be expected in physical grounds, it depends
explicitly on the nature of the created component (note the
prefactor $\rho + p=\gamma \rho$ on the RHS of (\ref{w18})).

What about the underlying physical
meaning of the new coupling term (\ref{w18})? Replacing (\ref{w10}) and
(\ref{w9}) in the time derivative of
(\ref{w13}) one finds
\begin{equation}
\dot\sigma \ = \ -\frac {1}{nT}
\biggl[u_{\mu}{T_{\phi}^{{\mu}{\nu}}}_{;\nu} + (\rho +
p)\Gamma_{\phi}\biggr].
\label{w19}
\end{equation}
Therefore, if (\ref{w18}) is valid, the specific entropy
of the created particles remains constant. A similar condition has
previously appeared in cosmologies with creation pressure\cite{LGA96},
as well as in cosmologies with a phenomenological time-dependent
$\Lambda$-term\cite{L96}, and was termed ``adiabatic" creation, because
the
entropy and the total number of particles increases but the ratio
$\sigma = {S \over N}$ remains constant. The condition $\dot\sigma = 0$
implies that the entropy $S$ of the thermalized bath increases
proportionally to the comoving number of created particles ($N \sim
nR^{3}$). More precisely, ${\dot S \over S} = {\dot N \over N} =
\Gamma_{\phi}$, which is a direct consequence of (\ref{w9}). Naturally,
even for ``adiabatic" decaying, there are out of equilibrium
contributions encoded in the temperature law (\ref{w17}). Without loss
of generality, let us analyze the case for a radiation bath. Inserting
the coupling term (\ref{w18}) into (\ref{w17}) and (\ref{w10}), and
rewriting (\ref{w9}) one finds the set of equations:
\begin{equation}
\label{w21}
{\dot T \over T} \ = \ -{\dot R \over R} \ + \ {\Gamma_{\phi} \over 3}
\quad,
\end{equation}
\begin{equation}
\label{w10A}
\dot \rho_{\phi} + 3H(\rho_{\phi} + p_{\phi} )= -{4 \over
3}\rho_{r}\Gamma_{\phi},
\end{equation}
\begin{equation}
\dot \rho_r +
4H\rho_{r} = {4 \over 3}\rho_{r} \Gamma_{\phi}\quad,
\label{w10B}
\end{equation}
\begin{equation}
{\dot n_r \over 3\,n_r\,H}  \ + \ 1 \ = \ {\Gamma_{\phi} \over 3H}\quad.
\label{w23}
\end{equation}
If the coupling term is negligible ($\Gamma_{\phi} \ll 3H$), we
see from (\ref{w21}) that the standard law is satisfied ($TR =
const$), whereas (\ref{w10B}) and (\ref{w23}) imply that $\rho_r
\propto R^{-4}$ and $n_r \propto R^{-3}$ as should be for an
adiabatic evolution. The opposite regime ($\Gamma_{\phi} \gg 3H$)
defines an extreme theoretical situation, where the decay process
is a phenomenon so fast that the dilution due to expansion is more
than compensated ($\dot T > 0$). An intermediary situation occurs
if this ratio is of the order of unity ($\Gamma_{\phi} \sim 3H$).
In this case, from (\ref{w21}) the bath temperature remains nearly
constant ($\dot T \sim 0$), and as a consistency check for the
whole thermodynamic scheme, we see directly from (\ref{w10B}) and
(\ref{w23}) that $n$ and $\rho$ are also nearly constant. Thus, a
warm scenario with a constant radiation energy density is
rigorously defined only when the scalar field decay rate is
fine-tuned to the inverse of the Hubble time ($\Gamma_{\phi} =
3H$). This is exactly what one should expect from physical
grounds: the dilution of the thermalized bath (due to expansion)
is balanced by the particle production regardless of the dynamic
details or any initial condition. Once this fine tuning is
assumed, the dynamic behavior of the universe, if exponential (de
Sitter), inflationary power law or even an ordinary decelerated
FRW type expansion is determined by the energetics of the system.
For example, if $V(\phi) \gg {\dot\phi}^{2} + \rho_r$, the
resulting scenario resembles the exponential warm inflation
originally suggested by Berera \cite{B95}. However, we see from
(\ref{w10B}) that the slow-roll conditions including the decay
rate will be modified. We recall that the standard slow-rollover
period requires $\Gamma \gg H$ so that $\rho_{r} \gg {\dot
\phi}^{2}$ (see Eqs.(13) and (17) of \cite{B95} and also (7) of
{\cite{TB20}). Within our approach these additional conditions are
not necessary, because the thermodynamic

constraint defining warm inflation is now reduced to $\Gamma_{\phi} =
3H$
regardless of the dynamic details. Such a condition also
provides a natural solution to the criticism of Yokoyama and Linde
\cite{YL99} for warm inflationary scenarios.

Although  an exponential warm inflation might be a viable scenario
within our approach, probably, the more realistic one is a power law
picture. It will happens, for instance, if the potential is proportional
to the kinetic term as recently suggested in some quintessence
cosmologies (see
\cite{JL99,kn:a2,W20} and Refs. therein). The simplest case is provided
by a scalar field obeying the equation of state, $p_{\phi} =
w\,\rho_{\phi}$,
where $w \in (-1,0)$ is a constant parameter. In this case, the
potential
$V_{\phi}$  and $\rho_{\phi}$ are proportional to  ${\dot{\phi}}^{2}$
\begin{equation}
V(\phi) \ = \ \frac{(1-w)}{2(1+w)}{\dot{\phi}}^{2} \ \  \ \mbox{and} \
\ \ \ \ \rho_{\phi} = \frac{1}{(1+w)}{\dot{\phi}}^{2}.
\label{w38}
\end{equation}
Using the new warm condition ($\Gamma_\phi = 3H, \rho_r =
\mbox{constant}$), the scalar field energy density is readily obtained
from (\ref{w10A}). One finds
\begin{equation}
\rho_{\phi} = \rho_{\phi_{i}}\,\left( \frac{R}{R_i}\right)^{-3(1+w)}\,
y\left( \frac{R}{R_i}\right),
\label{w25}
\end{equation}
with the quantity $y$ defined by
\begin{equation}
y\left( \frac{R}{R_i}\right)  \equiv  1  -
\frac{4\,\rho_{r}}{3(1+w)\rho_{\phi_{i}}}\left(
\frac{R}{R_i}\right)^{3(1+w)}
 +  \frac{4\,\rho_{r}}{3(1+w)\rho_{\phi_{i}}}.
\label{w26}
\end{equation}
Note that if $y = 0$ at a finite time $t^{*} > t_i$, equation
(\ref{w25})
yields $\rho_\phi(t^{*}) = 0$ showing that all the energy stored in the
scalar field has been  converted into photons. However, some of the
scalar field energy must still be retained at the end of the
inflationary process, in such a way
that ${\rho_\phi}(t_f) \sim \mbox{few} \rho_{r}$, where $t_f \ll t^{*}$.
In fact, the
end of inflation is defined by $\ddot R = 0$, which happens for
$\rho_{\phi_f} = -\frac{2}{1+3w}\,\rho_r$ ($w < -\frac{1}{3}$).
Naturally, the final of the
inflationary stage is a dynamic condition, and as such, it does not
means that the
scalar field decaying
process has already finished. This sort of coincidence may be only
artificially
adjusted so that some radiation entropy is presumably produced after
inflation.

Given the initial condition $\rho_{r} \ll \rho_{\phi_{i}}$
($\Omega_{\phi_i} \approx 1$), as long as the scalar field dominates,
the solution to the scale factor is
\begin{equation}
\frac{R}{R_{i}} \ \approx \ \left(
\frac{3(1+w)}{2\,H^{-1}_{i}}
\right)^{\frac{2}{3(1+w)}}(t - t_i)^{\frac{2}{3(1+w)}} \quad,
\label{w46}
\end{equation}
showing that $R(t)$
evolves like a power law, and a strict inflationary scenario may take
place for $w < -\frac{1}{3}$. The corresponding  number of ``e-folds",
$N =
\int_{t_i}^{t_f} \frac{\dot{R}}{R}\,dt$, is  given by
\begin{equation}
N = \frac{2}{3(1+w)}\,\ln\left( \frac{t_f}{t_i}\right)\quad,
\label{fol}
\end{equation}
and as shown in Table 1, it can be large enough to solve the main
cosmological problems.  In order to compute the expression for the
potential
$V(\phi)$, we first consider the evolution of $\phi$ with respect to
$R$.
Using (\ref{w10A}) and (\ref{w38}) one may see that the solution for the

scalar field is
\begin{equation}
\label{w48}
\phi \approx \phi_i  +  m_{pl}\sqrt\frac{3(1+w)}{8\pi}\,\ln\left(
\frac{R}{R_i}\right),
\end{equation}
where $H_i$ has been substituted by
$\sqrt{\frac{8\,\pi}{3\,m^{2}_{pl}}\rho_{\phi_i}}$.
Now, inserting  (\ref{w48}) and (\ref{w46}) into (\ref{w38}) we find
\begin{equation}
V(\phi) \approx  \frac{(1-w)}{2}\rho_{\phi_{i}}\mbox{exp}
\left(-\frac{\sqrt{24\pi(1+w)}}{m_{pl}}\,(\phi - \phi_i)\right),
\label{w49}
\end{equation}
\begin{equation}
\rho_{\phi} \ \approx \ \rho_{\phi_{i}}\,\mbox{exp}\left(
-\frac{\sqrt{24\pi (1+w)}}{m_{pl}}\,(\phi - \phi_i)\right),
\label{w50}
\end{equation}
which shows that for $w > -1$, the potential $V(\phi)$ is of the
exponential
type.

Now, we recall that the ratio $\frac{R_f}{R_i}$ is usually estimated
taking into account that the present entropy is $S_p \sim S_f \sim
10^{87}$, which holds if after $t_f$ the universe evolves adiabatically.
Thus, if
the decaying process continues until $t^{*} > t_{f}$ ($\rho_{\phi} \ll
\rho_r$), more entropy is produced which means that the entropy at $t_f$
should be somewhat smaller than $10^{87}$. In what follows we neglect
the entropy produced within this period. As one may check, if it is
included, the upper limits derived below on the values of
$\omega$ are relaxed.

By considering that the very early universe is radiation
dominated, it is a good approximation to set $\rho_{r} =
\frac{\pi^2}{30}\,g\,T^4$ and $S_i =
\frac{2\pi^2}{45}\,g\,T^3\,R_{i}^{3}$, where $g \sim 10^{2}$ is
the effective number of massless relativistic degrees of freedom.
Since $R_i = 2 t_i \sim 3\times 10^{-10}$ Gev$^{-1}$, where $t_i =
10^{-34}$ s, $T \sim 3\times 10^{13}$ Gev, we have $\rho_r \sim
2.7\times 10^{55}$ Gev$^{4}$ and $S_i \sim 3.2\times 10^{13}$.
>From $\frac{S_f}{S_i} = \left(\frac{R_f}{R_i}\right)^3$ one finds
$\frac{R_f}{R_i} \sim 10^{24}$, and inserting this value into
(\ref{w46}), it follows that
\begin{equation}
\Delta t \approx \frac{4}{3(1+w)} \times 10^{(36w + 2)}\, \mbox{s}\, ,
\label{w52}
\end{equation}
which yields the duration of the inflationary
phase in terms of $\omega$.  Recalling that when inflation ends the
field
energy density satisfy $\rho_{\phi_f} = -\frac{2}{1+3w}\,\rho_r$ and
expressing  $\rho_{\phi_f}$ as  $10^{m}\rho_r$, where $m$ is an
arbitrary
constant, one obtains from (\ref{w25})
\begin{equation}
\rho_{\phi_{i}} \sim \ \left( 10^m +
\frac{4}{3(1+w)}  \right)10^{72(1+w)}\,\rho_r ,
\label{w54}
\end{equation}
which means that $m >0$ only if $w < -\frac{1}{3}$.

By imposing the natural constraints
$\rho_{\phi_i} < \rho_{pl}$, where $\rho_{pl}\sim 10^{76}$ Gev$^4$ is
the
Planck  energy density, one obtains from (\ref{w54}) the inequality
$(10^m +
\frac{4}{3(1+w)}) <  10^{21 -72(1+w)}$, which provides the upper limit
$w <
-0.7$ for the quintessence  parameter. Therefore, one may conclude that
the
scalar field energy density in the  beginning of inflation satisfies the

required conditions, $\rho_{r} \ll \rho_{\phi_{i}} < \rho_{pl}$, only if
$w\in (-1,-0.7)$.

Table 1 display the limits on the main quantities in this power-law warm
picture. Note that the prescription $\rho_{r} \ll \rho_{\phi_{i}}$ is
always satisfied. The duration of inflation is also greater than in the
isentropic inflation. The number $N$ of ``e-folds" increases  with the
decreasing of $w$, and is always large enough to solve the
flatness-horizon problem.

Summarizing, in this letter we have derived a new interaction term
between the decaying inflaton field and a thermalized bath. This
coupling term is a direct consequence of thermodynamic arguments,
and it depends explicitly on the nature of the created particles,
now parameterized by their enthalpy density. In this way, a
consistent thermodynamic treatment defining clearly what we mean
by a sustained thermalized bath has emerged. In particular, if
$\Gamma_\phi \gg 3H$, its temperature increases in the course of
time regardless of the specific solution to $R(t)$. We recall that
such a constraint has been used as the key ingredient for warm
scenarios, however, as we have demonstrated, a weakened condition
($\Gamma_\phi \sim 3H$) is not only enough to define warm
inflation but holds regardless of the dynamic details. This means
that there is no physical preference for exponential or power-law
inflation since the evolution rate of the scale factor is now
completely independent of the warm condition. Once the decay width
of the scalar field and the volume expansion rate have been
adjusted, the evolution of the Universe is exclusively defined by
the energetic sector, that is, by the assumptions involving the
relative values of $V(\phi)$, ${\dot\phi^{2}}$ and $\rho_r =
\mbox{constant}$. This new viewpoint has been illustrated by a
simple warm power-law scenario, where the dynamics happens with no
slow-roll regime. The model depends on the equation of state,
$p_{\phi} = w\,\rho_{\phi}$, which fixes the form of the potential
$V(\phi)$ (exponential type). The duration of inflation is greater
than in the warm scenario \cite{B95}, and the model satisfies the
usual constraints if the cosmic parameter lies on the interval $-1
< w < -0.7$. For comparison, recent limits derived from a battery
of observational tests (including measurements from SNe type Ia),
constrain this parameter to be $\omega < -0.65$\cite{W20}.
Finally, we stress that our approach can also be applied for more
general models with scalar field coupled to a thermalized bath
(generic values of $\Gamma_{\phi}$). The extra bonus is the
complete control on all the thermodynamic aspects. These
thermodynamic results points to a new and somewhat more natural
way for implementing the basic idea of warm inflation, and, in
principle, should be justified from more fundamental calculations.

\section*{Acknowledgments}

The authors are grateful to Jackson M. F. Maia for reading the
manuscript and A. Berera for correspondence. This work was partially
supported by the Centro Latino Americano de F\'{\i}sica - CLAF and by
the Conselho Nacional de Desenvolvimento Cient\'{\i}fico e Tecnol\'ogico
- CNPq (Brazilian Research Agency).

\table
\begin{table}
\begin{center}
\begin{tabular}[t]{|c|c|c|c|c|}
\hline $w \ < $  & \ $\Delta t$ \ (sec.) \ & \ $\rho_{\phi_i}$ \
(Gev$^4$) \ & \
$\rho_{\phi_f}$  \ (Gev$^4$) \ & \ $N \ > $ \\   \hline  -0.71 &
$5.7\times
10^{-23}$ & $4.9 \times 10^{20}\,\rho_r$  & $1.77\,\rho_r$   & $ \ 59$
\\
\hline  -0.75 & $1.2 \times 10^{-25}$ & $6.9\times 10^{18}\,\rho_r$  &
$1.6\,\rho_r$   & $ \ 61$ \\  \hline
-0.8 & $1.1 \times 10^{-26}$& $2\times 10^{15}\,\rho_r$& $1.5\,\rho_r$
& $ \ 63$ \\ \hline
-0.9 & $5.3  \times 10^{-30}$& $2.3\times 10^{8}\,\rho_r$  &
$1.2\,\rho_r$ & $ \  72$ \\ \hline
\end{tabular}
\caption{Upper values for $w$ and other basic quantities. $\Delta t$,
$\rho_{\phi_i}$ ( $\rho_{\phi_f}$) and $N$ denote the duration of
inflation, the energy density of the scalar field in the begin (end) of
inflation, and the ``e-fold" number, respectively.}
\end{center}
\end{table}

\begin{thebibliography}{99}
\bibitem{N1} In order to avoid misunderstanding, we stress that
$\Gamma_{\phi}$ is different from the effective viscosity coefficient
$\Gamma$. These quantities have the same dimension, and, in general
$\Gamma$ is a function of $\Gamma_{\phi}$\cite{BGR,YL99}.
\end{thebibliography}
\end{document}